# Exploring causal physical mechanisms via non-gaussian linear models and deep kernel learning: applications for ferroelectric domain structures

Yongtao Liu,[1] Maxim Ziatdinov,[1,2] and Sergei V. Kalinin[1]

[1] Center for Nanophase Materials Sciences, Oak Ridge National Laboratory, Oak Ridge, TN 37831

[2] Computational Sciences and Engineering Division, Oak Ridge National Laboratory, Oak Ridge, Tennessee 37831, USA

Rapid emergence of the multimodal imaging in scanning probe, electron, and optical microscopies have brought forth the challenge of understanding the information contained in these complex data sets, targeting both the intrinsic correlations between different channels and further exploring the underpinning causal physical mechanisms. Here, we develop such analysis framework for the Piezoresponse Force Microscopy. We argue that under certain conditions, we can bootstrap experimental observations with the prior knowledge of materials structure to get information on certain non-observed properties, and demonstrate linear causal analysis for PFM observables. We further demonstrate that this approach can be extended to complex descriptors using the deep kernel learning (DKL) model. In this DKL analysis, we use the prior information on domain structure within the image to predict the physical properties. This analysis demonstrates the correlative relationships between morphology, piezoresponse, elastic property, etc. at nanoscale. The prediction of morphology and other physical parameters illustrates a mutual interaction between surface condition and physical properties in ferroelectric materials. This analysis is universal and can be extended to explore the correlative relationships of other multi-channel datasets.

Ferroelectric materials have remained a source of fascinating physical discovery through ~100 years history of the field.[1-2] The early discoveries of the piezoelectric, pyroelectric, and electrooptical effects enabled the broad applications of the ferroelectric single crystals and ceramics. Later, the presence of switchable polarization led to multiple concepts for ferroelectric-based data storage, including random access memories,[3] field-effect devices,[4-5] and ferroelectric tunneling devices.[6-7] Further opportunities emerged in the context of devices integrating ferroelectric and ferromagnetic components to enable multiferroic functionalities.[8-9]

The continuous progress in the field has brought forth the understanding of the crucial role of domain walls in emergent macroscopic functionalities. In particular, domain nucleation, wall motion, and pinning are directly responsible for the polarization switching phenomena, and are closely tied to dielectric properties of multidomain ferroelectrics.[10-12] The dynamics of the ferroelectric domain walls underpins the giant electromechanical responses in polycrystalline ceramics, morphotropic phase boundary materials,[13-17] and ferroelectric relaxors.[18-21] Interestingly, the domain structure of ferroelectric materials can be engineered via macroscopic poling conditions, allowing for the development of materials with enhanced properties.[22-23] Similarly, rapid progress in scattering techniques including the X-Ray and neutrons have enabled fundamental studies of these phenomena.[24-26]

Rapid progress in macroscopic studies of the polarization dynamics and its relationship with functional electromechanical, dielectric, and elastic properties have stimulated the interest in associated local mechanisms. While phase-field studies can be used to explore the domain structure evolution under the action of external fields,[27-28] realistic materials tend to contain defects and imperfections absent in phase-field models. Similarly, strong polarization-lattice pinning in ferroelectrics leading to Muller-Weinreich mechanisms for domain wall motion is difficult to represent via theoretical models.[29-32] In addition, the polarization dynamics on open ferroelectric surfaces is strongly affected by the ionic and internal screening.[33-34]

The development of scanning probe microscopy techniques including friction force microscopy, piezoresponse force microscopy (PFM),[35] and charge gradient microscopy[36] has opened a window into surface functionalities of ferroelectrics. The PFM allowed visualization of domain structure and its evolution under applied tip bias,[37] top electrode or lateral electrodes.[38-39] The band excitation (BE) PFM and acoustic force microscopy allowed to map the elastic properties,[40] whereas the charge collection microscopy have provided insight into the ionic screening.[41-42]

However, this emergence of large volumes of data in multiple information channels, in turn, necessitates understanding individual coupling mechanisms. Previously, we developed the *im2spec* and *spec2im* approach establishing the relationships between domain structure and hysteresis loops.[43] We have also explored the functional relationships between low-dimensional descriptors using Gaussian Process (GP) approach and postulated that GP uncertainty can be interpreted as sign of new physical phenomena.[44] Here, we extend this approach towards the exploration of the causal relationship between the surface parameters and establish the predictability of target functionalities from spatially distributed descriptors.

As a model system, we chosen a PbTiO$_3$ (PTO) thin film grown by metalorganic chemical vapor deposition (MOCVD) method on (001) KTaO$_3$ substrates with a SrRuO$_3$ buffer layer, as reported by H. Morioka et al.[45] Figure 1a shows the topography image of this PTO sample. Figure 1b-f show the corresponding band excitation (BE) PFM results. In BEPFM measurements, amplitude and phase vs. frequency curves were acquired at each spatial location, allowing for full probing of tip-surface interactions.[41, 46-47] Figure 1f shows several example amplitude vs. frequency curves from the locations marked on Figure 1b. These curves are fitted by a simple harmonic oscillator (SHO) model to extract the relevant parameters. Here, amplitude, resonance frequencies, and quality (Q)-factors are determined and are displayed as images. Shown in Figure 1b-c are the amplitude and phase images indicating the existence of both in-plane *a* domains and out-of-plane *c* domains. Shown in Figure 1d is the resonance frequency determined by tip-surface spring constants, which is proportional to the elastic property. Shown in Figure 1e is the Q-factor measuring the dissipative tip-surface interactions. In addition, the error for SHO fitting is a measure of nonlinearity.

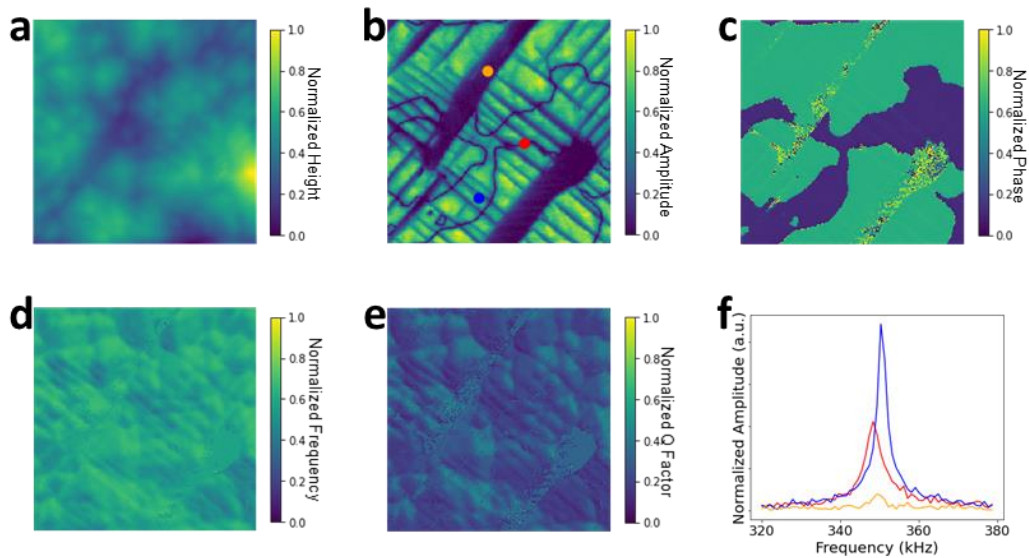

**Figure 1.** BE PFM of the PTO film. (a) Topography, (b) amplitude and (c) phase maps show c/c and a/c domains. (d) Resonance frequency map and (e) Q-factors maps. (f) Several examples of amplitude vs. frequency curves from the locations marked in (b).

To explore the relationship between the ferroelectric/ferroelastic domain structure and physical properties of the material, we seek to explore the causal relationships[48-52] between domain structure and parameters such as local elastic properties, energy dissipation, and nonlinearity. These parameters can be determined directly from the BE signals as illustrated above.

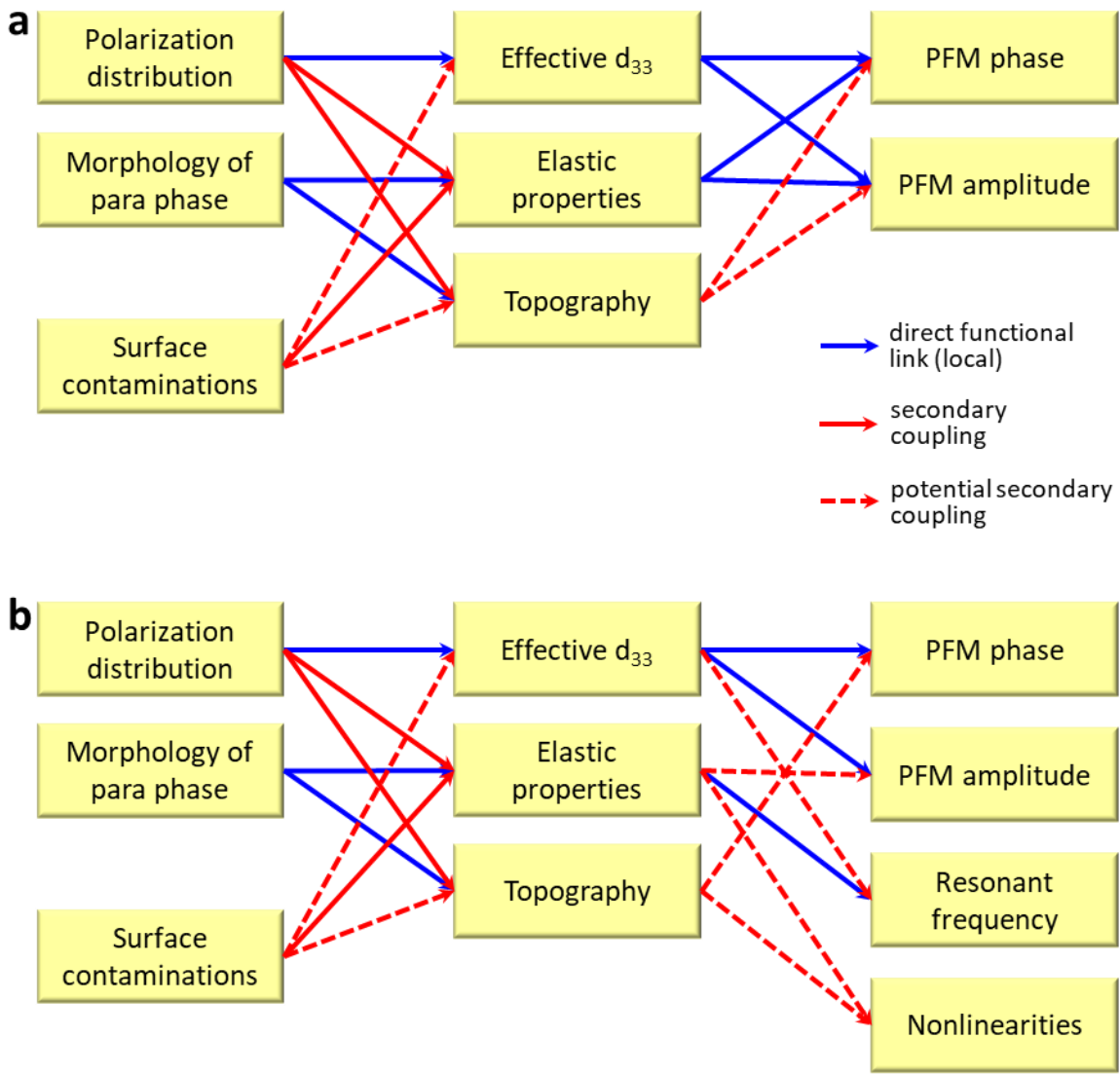

**Figure 2.** Causal relationships between intrinsic functionalities, ideal detected signals, and real detected signals for single frequency and band-excitation Piezoresponse Force Microscopy. Note that additional couplings such as polarization dependence of contaminant density can emerge, leading to the associated changes in observed signals. We also argue that under some conditions, we can use the observables as proxies for intrinsic functionalities (*causal bootstrapping*). For example, for material with a known crystallographic structure and high crystalline quality PFM image can be used as a proxy for polarization distribution.

Given that causal analysis is a relatively new development in physical sciences[53-54] and to date most of the relevant frameworks have been developed in the context of sociology, medicine, and economic sciences, here we briefly illustrate application of these concepts for physical systems. The general structural causal model approach has been developed by Pearl over 30 years,[49-50, 55] and provides the principled way of reconstructing the causal mechanisms encoded in the form of the directed acyclic graphs from observational and experimental data. Comparatively in physical sciences, the causal mechanisms are postulated, derived based on the theoretical considerations or from correlative observations. However, the discovery of causal structure from multimodal observational data is uncommon. We further note that physical sciences typically have a much more rigid structure of descriptors including the intrinsic materials parameters, observables, and coupling equations that comport to relevant physical mechanisms.

Here, we illustrate the causal relationships between intrinsic materials parameters, ideal functionalities, and measured signals for the single-frequency and band excitation PFM as shown in Figure 2. In both cases, the intrinsic physical descriptors are introduced as polarization distribution in the system (i.e. local domain state), morphology of the paraelectric phase (i.e. what the surface topography would have been if the material was not ferroelectric), and potential presence of surface contaminants.[56-57] For ideal PFM imaging, the observables include the effective electromechanical response $d_{33}$,[58-60] elastic properties of the tip-surface junction,[61-63] and surface topography. The relationship between these two groups of variables is defined by the physics of the PFM and SPM imaging. Here, the relationship between local polarization and $d_{33}$ is defined via contact mechanics of piezoelectric solid, and is the dominant coupling mechanism that can be analyzed via uncoupled[58, 64] and coupled[59] models. The relationship between the polarization and elastic properties of the material is defined via the secondary coupling that can be analyzed both via rigid continuum mechanic models[59-60, 65-66] and Ginzburg-Landau type models.[67] However, the contact resonance frequency is determined not only by effective Young's modulus, but also the surface morphology via the local radius of curvature.[63] Correspondingly, local elastic properties are determined both by morphology and polarization, with the former being the primary effect. Finally, the surface contaminates are known to strongly affect elastic properties, and can reduce PFM signal via the dielectric gap effect.[68] Note that this analysis excludes the mechanisms in which the polarization density is coupled to the contamination (or adsorbates)[69-73] via the formation of ferroionic phases.[74-75]

With the relationships between the materials descriptors and idealized measurables having been constructed, we can further analyze the relationship between ideal observables and real measurables. In a single-frequency PFM, the measured amplitude and phase are determined both by the surface topography and piezoresponse signal as the result of direct topographic cross-talk.[76-77] This coupling stems directly from the fundamental mechanics of the cantilever coupled to the surface and renders single frequency PFM in the vicinity of the resonance peak highly susceptible to topographic cross-talk. Comparatively, BE PFM allows to directly map the amplitude-frequency curve, allowing the resonance frequency and the amplitude to be separated and potentially allowing the contact non-linearities to be identified. Again, we note that the proposed diagram contains the physically determined causal links. At the same time, additional mechanisms such as non-linearities associated with polarization dynamics are possible.

Note that the diagrams in Figure 2 provide the physical mechanisms for single frequency and band excitation PFM. However, experimental observations are limited to the descriptors in the right column. However, here we pose that these observables can be used to bootstrap the material's intrinsic functionality. In this case, under the assumption that material has a constant composition, the observed PFM signal (right column) can be used as a proxy signal for intrinsic polarization distribution in the material (left column), allowing for causal analysis.

To explore this, we have first employed a linear non-gaussian acyclic model (LiNGAM) to construct the causal relationship of multiple channels in BEPFM dataset. LiNGAM was first proposed by Shimizu for the analysis of time series data.[52, 78-82] This model assumes that the relationships between the observed variables are linear, there are no non-observable confounders, and the external exogenous variables are non-Gaussian. In other words, the observations can be represented as:

$$\mathbf{z} = \mathbf{A}\mathbf{z} + \boldsymbol{\varepsilon} \tag{1}$$

where $\mathbf{z}$ is the vector of observations, $\mathbf{A}$ is the adjacency matrix establishing the relationships between them, and $\boldsymbol{\varepsilon}$ is the (non-Gaussian) noise vector. Under the assumption of the directed causal graph, the matrix, $\mathbf{A}$, has the lower diagonal shape. Correspondingly, LiNGAM seeks to find the linear decomposition of data via the independent component analysis (ICA)[83-84] approach. With the independent components established, the algorithm finds the optimal representation of the link matrix via permutations.

Shown in Figure 3a the LiNGAM causal chain and linear coefficient image based on normalized data with a directed path from morphology to morphology-curvature and no path between amplitude and phase. The variables in the LiNGAM causal map are generated from the BEPFM dataset, where the morphology corresponds to the topography and the morphology curvature is derived from morphology by calculating mean curvature. Generally, piezoresponse including amplitude and phase are found to be higher on the causal chain, elastic effect (i.e., frequency) being secondary, and surface geometry including morphology and morphology-curvature being the lower. Note that the LiNGAM without prior knowledge of directed path from morphology to morphology-curvature will lead to a path from morphology-curvature to morphology, as shown in Figure S1a, which is consistent with our known information (morphology-curvature is derived from morphology). In addition, by setting amplitude and phase as exogenous variables (in this case, we expect the piezoresponse is the major property and the primary driving force of this material's behavior), we obtained an identical causal map (Figure S1c) by setting no-path between amplitude and phase. The consideration of no-path between amplitude and phase is that both amplitude and phase represent piezoresponse but just different aspects, so we believe amplitude and phase should be parallel in causal chain. Moreover, the LiNGAM analysis is also performed on the normalized data (Figure S2), which results in morphology being a higher level in causal map and the frequency still being a lower level. This can be understood as the role of surface and interface energy in affecting the elasticity of this material.

Based on these analyses, we established the relationship between the local observables and the material's properties they represent on a single point level. However, this approach necessitates the creation of a large number of descriptors, including the gradients and curvature for topography, and potentially equivalent descriptors for the polarization. At the same time, of interest are the relationships between the observables represented by multidimensional descriptors such as local polarization and topographic patches. While these can be expected to contain information on gradients, curvature, as well as more complex descriptors, it is not clear ad hoc which of these parameters are relevant. The simple dimensionality reduction methods, while in principle possible, will still generate a large number of potentially spurious descriptors.

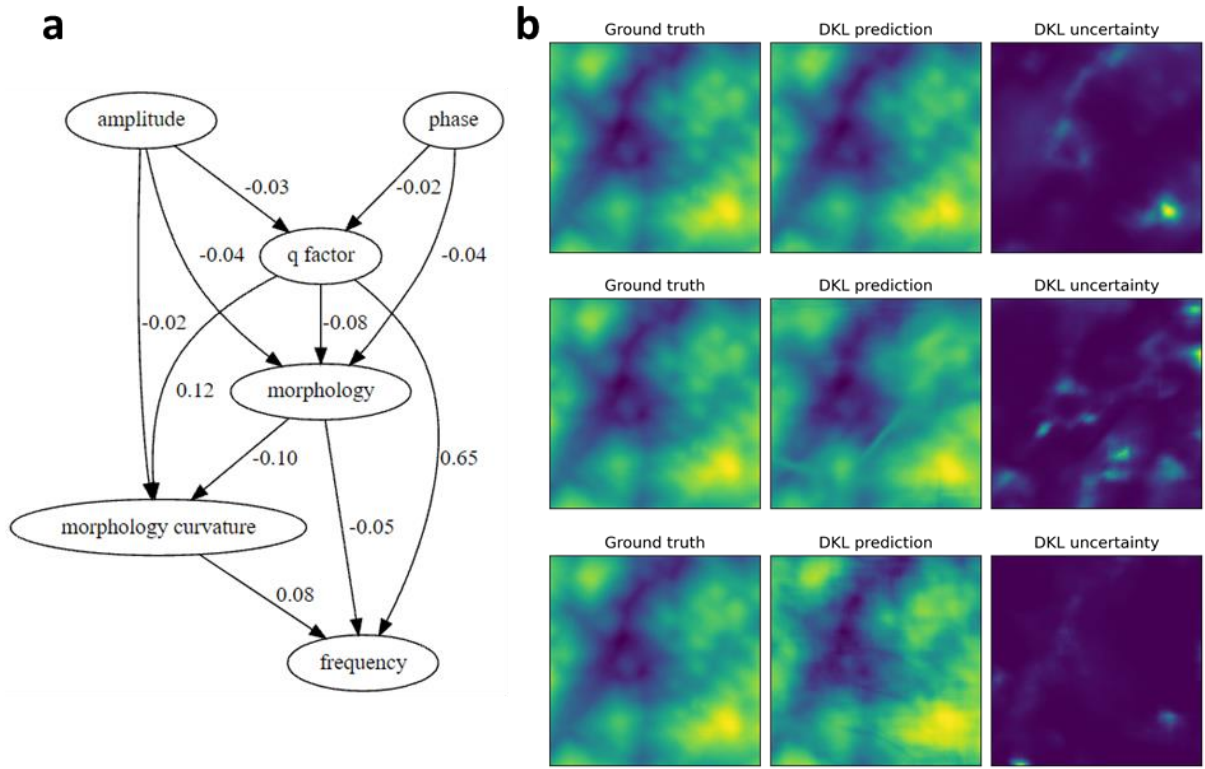

**Figure 3. Causal and DKL analysis. a)** Causal chain and linear coefficient map by LiNGAM analysis with directed paths from morphology to morphology-curvature and no-paths between amplitude and phase. **b)** DKL prediction of morphology (topography) based on amplitude (top), phase (middle), and frequency (bottom). The DKL prediction and uncertainty are computed as mean and variance of the samples in Eq. 4.

To explore these relationships practically, we adapted a deep kernel learning (DKL)[85] approach where a neural network converts high-dimensional input data into a set of low-dimensional descriptors on which a standard GP kernel operates. The DKL regression model for the inputs $X = (x_1, ..., x_N)$ and targets $\mathbf{y} = (y_1, ..., y_N)$ can be defined the same way as a classical GP regression model, that is,

$$y_n = f(x_n) + \varepsilon_n, \tag{1}$$

$$f \sim \mathcal{GP}(0, k_{DKL}), \tag{2}$$

where $\varepsilon_n$ represents normally distributed noise with zero mean and variance $s_n^2$. The $k_{DKL}$ is a deep kernel function combining the feedforward neural network ("feature extractor") and a standard GP kernel, here chosen as radial basis function (RBF),

$$k_{DKL}(x, x'|\mathbf{w}, \boldsymbol{\theta}) = k_{RBF}(g(x|\mathbf{w}), g(x'|\mathbf{w})|\boldsymbol{\theta}) \qquad (3)$$

where $\mathbf{w}$ are the weights of the neural network $g$, and $\boldsymbol{\theta}$ are the hyperparameters (amplitude and length scale) of the RBF 'base' kernel. These parameters, together with the regression model noise $s_n^2$, are learned simultaneously by maximizing the log evidence via a stochastic variational inference.[86] The DKL predictions for a set of new/test points $X_*$ are computed by sampling from the multivariate normal (*MVN*) posterior as

$$\mathbf{f}_* \sim MVN(\mu^{\text{post}}, \Sigma^{\text{post}}) \qquad (4)$$

$$\mu^{\text{post}} = K(X_*, X)(K(X, X) + s_n^2 I)^{-1} \mathbf{y} \qquad (5)$$

$$\Sigma^{\text{post}} = K(X_*, X_*) - K(X_*, X)(K(X, X) + s_n^2 I)^{-1} K(X, X_*) \qquad (6)$$

where $K_{ij} = k_{DKL}(x_i, x_j)$. Since in the DKL approach the GP operates in the latent (embedding) space learned by a neural network from the high-dimensional data, the overall structure can be understood as the data-informed (deep) kernel.

To explore the relationship between BEPFM channels, we generated image patches with domain structures from the full BEPFM images. As shown in Figure 4a-b, Figure 4a shows a full amplitude image and Figure 4b shows example small image patches (including amplitude, phase, morphology, and frequency) extracted from the region marked on Figure 4a. These image patches illustrating known domain structure are used as input in DKL analysis (shown in Figure 4c-d), the output of DKL is the property corresponding to these structures. In DKL train phase (Figure 4c), we feed the model with known structure and property, while in the prediction phase we only feed the model with known structure (Figure 4d) and allow the model to predict the corresponding property according to Eq. (4-6).

Based on the causal chain map, we performed DKL analysis to predict the morphology (lowest in causal map) from piezo properties and elastic property. Shown in Figure 3b is the DKL prediction of morphology based on amplitude (top), phase (middle), and frequency (bottom). All DKL predictions show a good consistency with the ground truth morphology. DKL uncertainty maps show that high variance mostly located in the large in-plane *a* domain. Note that the causal chain map in Figure 3a is constructed from normalized data; the LiNGAM analysis on unnormalized data will result in a different causal chain map, as shown in Figure S1, where the morphology is sorted out from other physical properties. Nonetheless, amplitude and phase are still higher on the causal chain and frequency and Q-factor being lower.

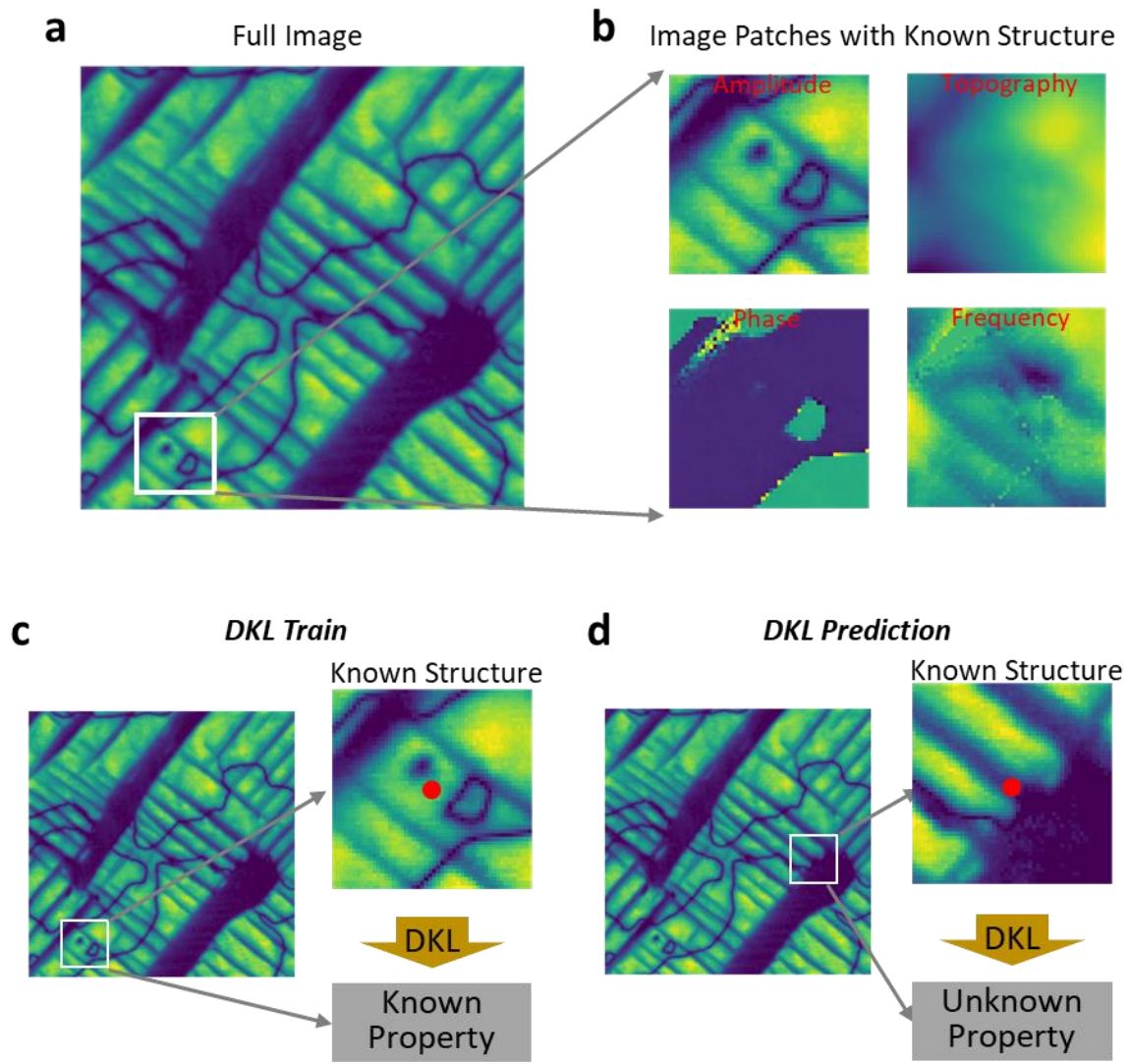

**Figure 4. Data processing and DKL analysis process. a), b),** show the extraction of image patches from a full image; **b),** shows examples of image patch for amplitude, phase, topography, and frequency that include structural information. **c), d),** illustrates the DKL training and prediction processes based on the structural images.

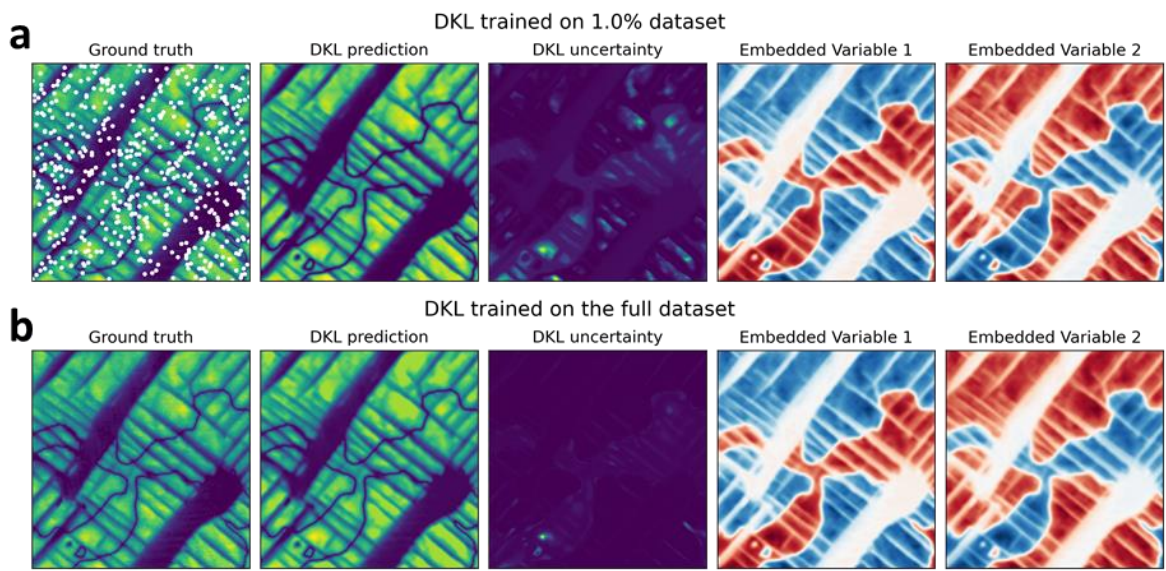

**Figure 5. DKL analysis of amplitude based on polarization map**. **(a)** DKL prediction based on random 1.0 % of data and corresponding embedded latent variables, with the white dots in ground truth amplitude data indicating the data used for DKL training **(b),** DKL prediction based on the full data and corresponding embedded latent variables, these latent variables show large domains (right) and fine domains (left), respectively.

To further illustrate this approach, we perform DKL analysis to predict the amplitude based on the polarization map. Shown in Figure 5a is the DKL prediction based on 1 % of the polarization data, where the data used for DKL analysis is marked as white dots in the ground truth amplitude in Figure 5a. The DKL prediction reconstructed the amplitude generally well when comparing with the ground truth. The DKL uncertainty map indicates higher variances in in-plane *a* domains, which is reasonable as the in-plane polarization is not known in a vertical PFM dataset. The corresponding embedded latent variables shown in Figure 5a also illustrated nice domain structures consistent with those in the ground truth. Figure 5b shows the DKL prediction based on full polarization data. In this case, DKL reconstructed the amplitude very well and the uncertainty map shows mostly zero values everywhere. We also systematically investigated the performance of DKL prediction as a function of training data size, which are shown in Figure S3-S11. The predictive performance of DKL is represented as the mean squared error (MSE) between prediction and ground truth, with smaller MSE indicating better performance. As seen in Figure S3-S11, generally, < 10 % of data for DKL training results in a good reconstruction of the full image.

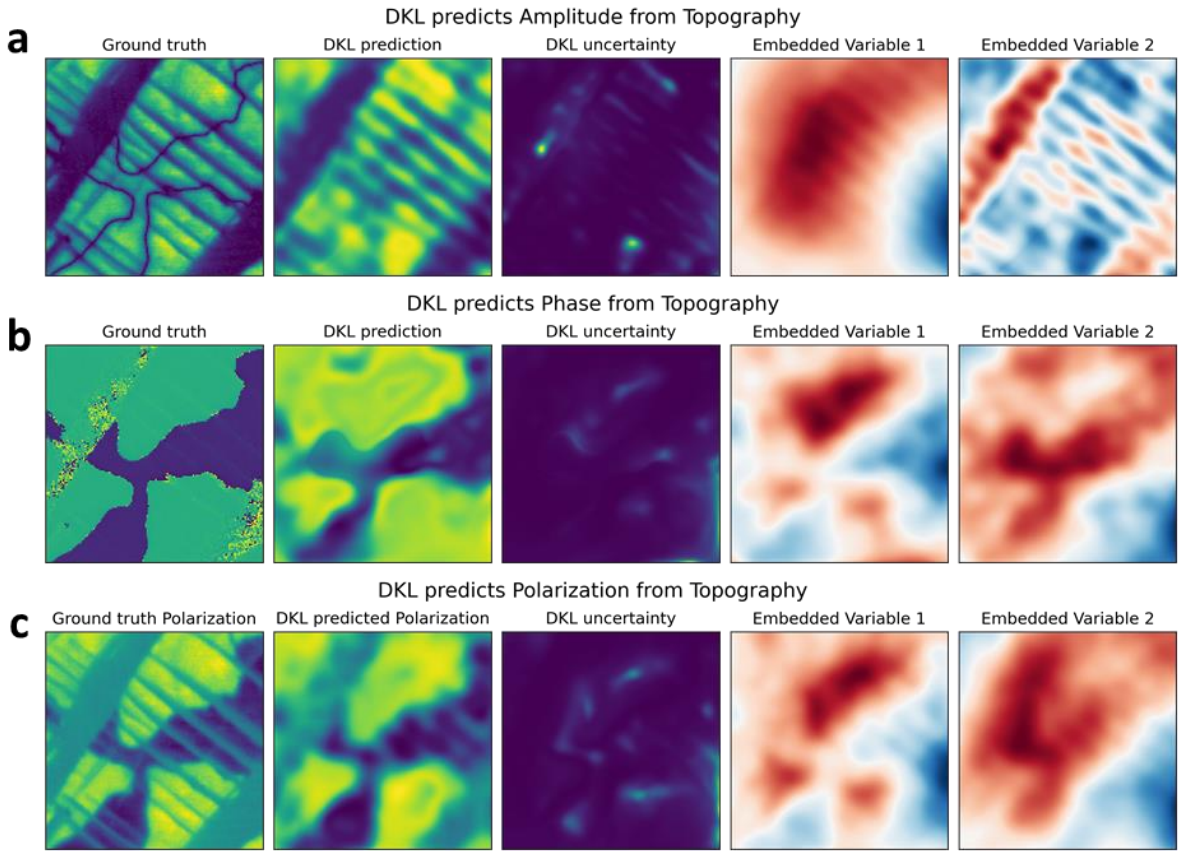

**Figure 6: DKL prediction of piezoresponse based on morphology in comparison with ground truth, and corresponding DKL uncertainties and embedded latent variables. a),** DKL prediction of amplitude. **b)** DKL prediction of phase. **c)** DKL prediction of polarization.

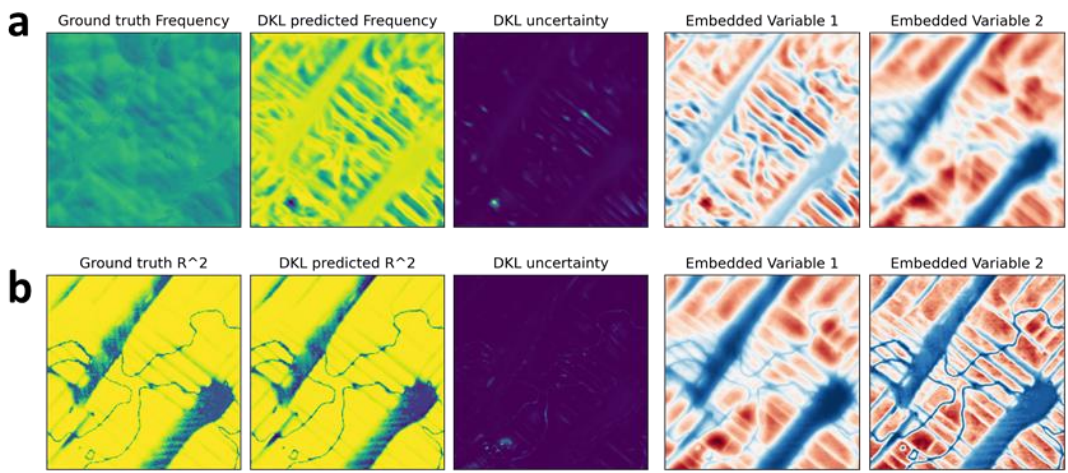

**Figure 7. DKL prediction of elasticity and nonlinearity based on polarization in comparison with ground truth, and corresponding DKL uncertainties and embedded latent variables. a),**

DKL prediction of elasticity that is reflected in resonance frequency map. **b)** DKL prediction of nonlinearity that is reflected in the SHO fitting $R^2$ map.

This approach is further extended to predict piezoresponse from morphology. In this case, large topographic patch size results in better reconstruction of piezoresponse. The results shown in Figure 6 is the prediction of amplitude, phase, and polarization from topography data with a patch size of 100*100. Shown in Figure 6a, the prediction of amplitude demonstrates a good reconstruction of a/c domains with distinct amplitude contrast in the vertical PFM. In the meantime, the DKL embedded latent variables illustrate the topographic background (compare embedded variable 1 with the topography in Figure 1a) and the a/c domain structures (embedded variable 2). In Figure 6b, the prediction of phase interestingly reconstructs the c/c domain contrast. It is generally believed that a/c ferroelastic domains are tied with topographic geometry, while this result (Figure 6b) indicates that there is also a potential relationship between c/c ferroelectric domains and topographic geometry. Embedded variables in Figure 6b also illustrate the topographic background (embedded variable 2) and the c/c domain structures. The prediction of polarization in Figure 6c is a combination of amplitude and phase, which reconstructs both a/c and c/c domains. Generally, in these DKL predictions (Figure 6), the embedded variables disentangle the pure topographic background and domain structures.

Moreover, this approach can be extended to predict elastic property and nonlinearity from piezoresponse. Shown in Figure 7a is the DKL prediction of resonance frequency from polarization, the DKL prediction mostly reconstructed the elastic variation over the image. Shown in Figure 7b is the DKL prediction of $R^2$ (R-squared for SHO fitting) of SHO fitting that is related to the nonlinearity, where the DKL prediction shows a good consistency with the ground truth. In both cases here, the embedded variables are almost identical illustrating the domain structures.

In summary, we utilized the non-gaussian linear models and deep kernel learning to analyze the causal relationships between the physical parameters in multi-channels BEPFM datasets. BEPFM offers diverse information including morphology, piezoresponse, elasticity, and nonlinearity at nanoscale levels. Our analyses using DKL demonstrated the correlation between these parameters. For example, the DKL prediction of piezoresponse from morphology and the prediction of morphology from piezoresponse indicate the mutual correlation between morphology and piezoresponse. It is generally believed that the a/c ferroelastic domains results in orientation contrast in morphology, while the effect of c/c ferroelectric domains on morphology is rarely considered. However, the DKL prediction of c/c domain structures from morphology reveals a potential interaction between c/c domains and surface geometry. This approach can also be extended to explore the physical correlations of other multi-channels datasets.

## Methods:

*Data analysis*

The detailed methodologies of DKL analysis on BEPFM dataset are established in Jupyter notebooks and are available from https: https://git.io/Jwnqy.

*$PbTiO_3$ sample*

The *$PbTiO_3$* film was grown by chemical vapor deposition on a $SrRuO_3$ bottom electrode on a $KTaO_3$ substrate.

*BEPFM measurement*

The PFM was performed using an Oxford Instrument Asylum Research Cypher microscope with Budget Sensor Multi75E-G Cr/Pt coated AFM probes (~3 N/m). Band excitation data are acquired with a National Instruments DAQ card and chassis operated with a LabView framework.

## Conflict of Interest

The authors declare no conflict of interest.

## Authors Contribution

S.V.K. conceived the project and M.Z. realized the DKL workflow. Y.L. performed detailed analyses with basic workflow from M.Z. All authors contributed to discussions and the final manuscript.

## Acknowledgements

This effort (ML) was supported as part of the center for 3D Ferroelectric Microelectronics (3DFeM), an Energy Frontier Research Center funded by the U.S. Department of Energy (DOE), Office of Science, Basic Energy Sciences under Award Number DE-SC0021118 (Y.L., S.V.K.), and the Oak Ridge National Laboratory's Center for Nanophase Materials Sciences (CNMS), a U.S. Department of Energy, Office of Science User Facility (M.Z). The authors gratefully acknowledge Prof. Hiroshi Funakubo for the PTO samples used in this work.

## Data Availability Statement

The data that support the findings of this study are available at https: https://git.io/Jwnqy.

# References

1. Whatmore, R. W.; You, Y.-M.; Xiong, R.-G.; Eom, C.-B., 100 years of ferroelectricity—A celebration. AIP Publishing LLC: 2021.
2. Bhalla, A. S.; Saxena, A., Ferroelectricity: 100 years on. *Physics World* **2021,** *33* (11), 38.
3. Dearaujo, C. A. P.; Cuchiaro, J. D.; McMillan, L. D.; Scott, M. C.; Scott, J. F., FATIGUE-FREE FERROELECTRIC CAPACITORS WITH PLATINUM-ELECTRODES. *Nature* **1995,** *374* (6523), 627-629.
4. Miller, S. L.; McWhorter, P. J., PHYSICS OF THE FERROELECTRIC NONVOLATILE MEMORY FIELD-EFFECT TRANSISTOR. *J. Appl. Phys.* **1992,** *72* (12), 5999-6010.
5. Mathews, S.; Ramesh, R.; Venkatesan, T.; Benedetto, J., Ferroelectric field effect transistor based on epitaxial perovskite heterostructures. *Science* **1997,** *276* (5310), 238-240.
6. Maksymovych, P.; Jesse, S.; Yu, P.; Ramesh, R.; Baddorf, A. P.; Kalinin, S. V., Polarization Control of Electron Tunneling into Ferroelectric Surfaces. *Science* **2009,** *324* (5933), 1421-1425.
7. Tsymbal, E. Y.; Kohlstedt, H., Applied physics - Tunneling across a ferroelectric. *Science* **2006,** *313* (5784), 181-183.
8. Bocher, L.; Gloter, A.; Crassous, A.; Garcia, V.; March, K.; Zobelli, A.; Valencia, S.; Enouz-Vedrenne, S.; Moya, X.; Marthur, N. D.; Deranlot, C.; Fusil, S.; Bouzehouane, K.; Bibes, M.; Barthelemy, A.; Colliex, C.; Stephan, O., Atomic and Electronic Structure of the BaTiO3/Fe Interface in Multiferroic Tunnel Junctions. *Nano Lett.* **2012,** *12* (1), 376-382.
9. Jayachandran, K. P.; Guedes, J. M.; Rodrigues, H. C., Solutions for maximum coupling in multiferroic magnetoelectric composites by material design. *Sci Rep* **2018,** *8*, 9.
10. Damjanovic, D., Ferroelectric, dielectric and piezoelectric properties of ferroelectric thin films and ceramics. *Rep. Prog. Phys.* **1998,** *61* (9), 1267-1324.
11. Setter, N.; Damjanovic, D.; Eng, L.; Fox, G.; Gevorgian, S.; Hong, S.; Kingon, A.; Kohlstedt, H.; Park, N. Y.; Stephenson, G. B.; Stolitchnov, I.; Tagantsev, A. K.; Taylor, D. V.; Yamada, T.; Streiffer, S., Ferroelectric thin films: Review of materials, properties, and applications. *J. Appl. Phys.* **2006,** *100* (5).
12. Bintachitt, P.; Jesse, S.; Damjanovic, D.; Han, Y.; Reaney, I. M.; Trolier-McKinstry, S.; Kalinin, S. V., Collective dynamics underpins Rayleigh behavior in disordered polycrystalline ferroelectrics. *Proc. Natl. Acad. Sci. U. S. A.* **2010,** *107* (16), 7219-7224.
13. Grinberg, I.; Suchomel, M. R.; Davies, P. K.; Rappe, A. M., Predicting morphotropic phase boundary locations and transition temperatures in Pb- and Bi-based perovskite solid solutions from crystal chemical data and first-principles calculations. *J. Appl. Phys* **2005,** *98* (9).
14. Iwata, M.; Katsuraya, K.; Suzuki, I.; Maeda, M.; Yasuda, N.; Ishibashi, Y., Domain wall observation and dielectric anisotropy in PZN-PT by SPM. *Materials Science and Engineering B-Solid State Materials for Advanced Technology* **2005,** *120* (1-3), 88-90.
15. Woodward, D. I.; Knudsen, J.; Reaney, I. M., Review of crystal and domain structures in the $PbZr_xTi_{1-x}O_3$ solid solution. *Phys. Rev. B* **2005,** *72* (10).
16. Zeches, R. J.; Rossell, M. D.; Zhang, J. X.; Hatt, A. J.; He, Q.; Yang, C. H.; Kumar, A.; Wang, C. H.; Melville, A.; Adamo, C.; Sheng, G.; Chu, Y. H.; Ihlefeld, J. F.; Erni, R.; Ederer, C.; Gopalan, V.; Chen, L. Q.; Schlom, D. G.; Spaldin, N. A.; Martin, L. W.; Ramesh, R., A Strain-Driven Morphotropic Phase Boundary in $BiFeO(3)$. *Science* **2009,** *326* (5955), 977-980.
17. Rodel, J.; Jo, W.; Seifert, K. T. P.; Anton, E. M.; Granzow, T.; Damjanovic, D., Perspective on the Development of Lead-free Piezoceramics. *J. Am. Ceram. Soc.* **2009,** *92* (6), 1153-1177.
18. Cross, L. E., RELAXOR FERROELECTRICS. *Ferroelectrics* **1987,** *76* (3-4), 241-267.

19.     Westphal, V.; Kleemann, W.; Glinchuk, M. D., DIFFUSE PHASE-TRANSITIONS AND RANDOM-FIELD-INDUCED DOMAIN STATES OF THE RELAXOR FERROELECTRIC PBMG1/3NB2/3O3. *Phys. Rev. Lett.* **1992,** *68* (6), 847-850.

20.     Xu, Z.; Kim, M. C.; Li, J. F.; Viehland, D., Observation of a sequence of domain-like states with increasing disorder in ferroelectrics. *Philos. Mag. A-Phys. Condens. Matter Struct. Defect Mech. Prop.* **1996,** *74* (2), 395-406.

21.     Kalinin, S. V.; Rodriguez, B. J.; Budai, J. D.; Jesse, S.; Morozovska, A. N.; Bokov, A. A.; Ye, Z. G., Direct evidence of mesoscopic dynamic heterogeneities at the surfaces of ergodic ferroelectric relaxors. *Phys. Rev. B* **2010,** *81* (6), 064107.

22.     Wada, S.; Yako, K.; Kakemoto, H.; Tsurumi, T.; Kiguchi, T., Enhanced piezoelectric properties of barium titanate single crystals with different engineered-domain sizes. *Journal of Applied Physics* **2005,** *98* (1), 014109.

23.     Rojac, T.; Damjanovic, D., Domain walls and defects in ferroelectric materials. *Japanese Journal of Applied Physics* **2017,** *56* (10S), 10PA01.

24.     Glaum, J.; Granzow, T.; Rödel, J., Evaluation of domain wall motion in bipolar fatigued lead-zirconate-titanate: A study on reversible and irreversible contributions. *Journal of Applied Physics* **2010,** *107* (10), 104119.

25.     Jones, J. L.; Slamovich, E. B.; Bowman, K. J., Domain texture distributions in tetragonal lead zirconate titanate by x-ray and neutron diffraction. *Journal of Applied Physics* **2005,** *97* (3), 034113.

26.     Jones, J. L.; Hoffman, M.; Daniels, J. E.; Studer, A. J., Direct measurement of the domain switching contribution to the dynamic piezoelectric response in ferroelectric ceramics. *Applied physics letters* **2006,** *89* (9), 092901.

27.     Chen, L. Q., Phase-field models for microstructure evolution. *Annual Review of Materials Research* **2002,** *32*, 113-140.

28.     Choudhury, S.; Zhang, J. X.; Li, Y. L.; Chen, L. Q.; Jia, Q. X.; Kalinin, S. V., Effect of ferroelastic twin walls on local polarization switching: Phase-field modeling. *Appl. Phys. Lett.* **2008,** *93* (16).

29.     Burtsev, E. V.; Chervonobrodov, S. P., SOME PROBLEMS OF 180-DEGREES-SWITCHING IN FERROELECTRICS. *Ferroelectrics* **1982,** *45* (1-2), 97-106.

30.     Burtsev, E. V.; Chervonobrodov, S. P., A TENTATIVE MODEL FOR DESCRIBING THE POLARIZATION REVERSAL PROCESS IN FERROELECTRICS IN WEAK ELECTRIC-FIELDS. *Kristallografiya* **1982,** *27* (5), 843-850.

31.     Burtsev, E. V.; Chervonobrodov, S. P., SOME PECULIARITIES OF 180-DEGREES-DOMAIN WALL MOTION IN THIN FERROELECTRIC-CRYSTALS. *Ferroelectr. Lett. Sect.* **1983,** *44* (10), 293-299.

32.     Miller, R. C.; Weinreich, G., MECHANISM FOR THE SIDEWISE MOTION OF 180-DEGREES DOMAIN WALLS IN BARIUM TITANATE. *Physical Review* **1960,** *117* (6), 1460-1466.

33.     Morozovska, A. N.; Eliseev, E. A.; Kurchak, A. I.; Morozovsky, N. V.; Vasudevan, R. K.; Strikha, M. V.; Kalinin, S. V., Effect of surface ionic screening on the polarization reversal scenario in ferroelectric thin films: Crossover from ferroionic to antiferroionic states. *Physical Review B* **2017,** *96* (24), 245405.

34.     Kalinin, S. V.; Kim, Y.; Fong, D. D.; Morozovska, A. N., Surface-screening mechanisms in ferroelectric thin films and their effect on polarization dynamics and domain structures. *Reports on Progress in Physics* **2018,** *81* (3), 036502.

35.     Kolosov, O.; Gruverman, A.; Hatano, J.; Takahashi, K.; Tokumoto, H., NANOSCALE VISUALIZATION AND CONTROL OF FERROELECTRIC DOMAINS BY ATOMIC-FORCE MICROSCOPY. *Phys. Rev. Lett.* **1995,** *74* (21), 4309-4312.

36.     Hong, S.; Tong, S.; Park, W. I.; Hiranaga, Y.; Cho, Y.; Roelofs, A., Charge gradient microscopy. *Proceedings of the National Academy of Sciences* **2014,** *111* (18), 6566-6569.

37. Vasudevan, R.; Jesse, S.; Kim, Y.; Kumar, A.; Kalinin, S., Spectroscopic imaging in piezoresponse force microscopy: New opportunities for studying polarization dynamics in ferroelectrics and multiferroics. *MRS communications* **2012,** *2* (3), 61-73.
38. Balke, N.; Gajek, M.; Tagantsev, A. K.; Martin, L. W.; Chu, Y. H.; Ramesh, R.; Kalinin, S. V., Direct Observation of Capacitor Switching Using Planar Electrodes. *Adv. Funct. Mater.* **2010,** *20* (20), 3466-3475.
39. Gruverman, A.; Alexe, M.; Meier, D., Piezoresponse force microscopy and nanoferroic phenomena. *Nature communications* **2019,** *10* (1), 1-9.
40. Lindgren, G.; Ievlev, A.; Jesse, S.; Ovchinnikova, O. S.; Kalinin, S. V.; Vasudevan, R. K.; Canalias, C., Elasticity Modulation Due to Polarization Reversal and Ionic Motion in the Ferroelectric Superionic Conductor KTiOPO4. *Acs Applied Materials & Interfaces* **2018,** *10* (38), 32298-32303.
41. Jesse, S.; Kalinin, S. V.; Proksch, R.; Baddorf, A.; Rodriguez, B., The band excitation method in scanning probe microscopy for rapid mapping of energy dissipation on the nanoscale. *Nanotechnology* **2007,** *18* (43), 435503.
42. Griggio, F.; Jesse, S.; Kumar, A.; Marincel, D.; Tinberg, D.; Kalinin, S.; Trolier-McKinstry, S., Mapping piezoelectric nonlinearity in the Rayleigh regime using band excitation piezoresponse force microscopy. *Applied Physics Letters* **2011,** *98* (21), 212901.
43. Ziatdinov, M.; Ghosh, A.; Wong, T.; Kalinin, S. V., AtomAI: A Deep Learning Framework for Analysis of Image and Spectroscopy Data in (Scanning) Transmission Electron Microscopy and Beyond. *arXiv preprint arXiv:2105.07485* **2021**.
44. Ziatdinov, M.; Creange, N.; Zhang, X.; Morozovska, A.; Eliseev, E.; Vasudevan, R. K.; Takeuchi, I.; Nelson, C.; Kalinin, S. V., Predictability as a probe of manifest and latent physics: The case of atomic scale structural, chemical, and polarization behaviors in multiferroic Sm-doped BiFeO3. *Appl. Phys. Rev.* **2021,** *8* (1), 011403.
45. Morioka, H.; Yamada, T.; Tagantsev, A. K.; Ikariyama, R.; Nagasaki, T.; Kurosawa, T.; Funakubo, H., Suppressed polar distortion with enhanced Curie temperature in in-plane 90-domain structure of a-axis oriented PbTiO3 Film. *Applied Physics Letters* **2015,** *106* (4), 042905.
46. Jesse, S.; Kalinin, S. V., Band excitation in scanning probe microscopy: sines of change. *J. Phys. D-Appl. Phys.* **2011,** *44* (46), 464006.
47. Jesse, S.; Vasudevan, R. K.; Collins, L.; Strelcov, E.; Okatan, M. B.; Belianinov, A.; Baddorf, A. P.; Proksch, R.; Kalinin, S. V., Band Excitation in Scanning Probe Microscopy: Recognition and Functional Imaging. *Annual Review of Physical Chemistry, Vol 65* **2014,** *65*, 519-536.
48. Bareinboim, E.; Pearl, J., Causal inference and the data-fusion problem. *Proceedings of the National Academy of Sciences of the United States of America* **2016,** *113* (27), 7345-7352.
49. Pearl, J., The Seven Tools of Causal Inference, with Reflections on Machine Learning. *Commun. ACM* **2019,** *62* (3), 54-60.
50. Pearl, J., A Linear "Microscope" for Interventions and Counterfactuals. *J. Causal Inference* **2017,** *5* (1), 15.
51. Peters, J.; Mooij, J. M.; Janzing, D.; Scholkopf, B., Causal Discovery with Continuous Additive Noise Models. *J. Mach. Learn. Res.* **2014,** *15*, 2009-2053.
52. Shimizu, S.; Hoyer, P. O.; Hyvarinen, A.; Kerminen, A., A linear non-Gaussian acyclic model for causal discovery. *J. Mach. Learn. Res.* **2006,** *7*, 2003-2030.
53. Ziatdinov, M.; Nelson, C. T.; Zhang, X. H.; Vasudevan, R. K.; Eliseev, E.; Morozovska, A. N.; Takeuchi, I.; Kalinin, S. V., Causal analysis of competing atomistic mechanisms in ferroelectric materials from high-resolution scanning transmission electron microscopy data. *npj Comput. Mater.* **2020,** *6* (1), 9.
54. Vasudevan, R. K.; Ziatdinov, M.; Vlcek, L.; Kalinin, S. V., Off-the-shelf deep learning is not enough, and requires parsimony, Bayesianity, and causality. *npj Comput. Mater.* **2021,** *7* (1), 6.
55. Pearl, J., On the Interpretation of do(x). *J. Causal Inference* **2019,** *7* (1), 6.

56. Ievlev, A. V.; Brown, C. C.; Agar, J. C.; Velarde, G. A.; Martin, L. W.; Belianinov, A.; Maksymovych, P.; Kalinin, S. V.; Ovchinnikova, O. S., Nanoscale Electrochemical Phenomena of Polarization Switching in Ferroelectrics. *Acs Applied Materials & Interfaces* **2018,** *10* (44), 38217-38222.
57. Ievlev, A. V.; Morozovska, A. N.; Eliseev, E. A.; Shur, V. Y.; Kalinin, S. V., Ionic field effect and memristive phenomena in single-point ferroelectric domain switching. *Nat. Commun.* **2014,** *5*.
58. Eliseev, E. A.; Kalinin, S. V.; Jesse, S.; Bravina, S. L.; Morozovska, A. N., Electromechanical detection in scanning probe microscopy: Tip models and materials contrast. *J. Appl. Phys* **2007,** *102* (1).
59. Kalinin, S. V.; Karapetian, E.; Kachanov, M., Nanoelectromechanics of piezoresponse force microscopy. *Phys. Rev. B* **2004,** *70* (18), 184101.
60. Kalinin, S. V.; Shin, J.; Kachanov, M.; Karapetian, E.; Baddorf, A. P., Nanoelectromechanics of piezoresponse force microscopy: Contact properties, fields below the surface and polarization switching. In *Ferroelectric Thin Films Xii*, HoffmannEifert, S.; Funakubo, H.; Joshi, V.; Kingon, A. I.; Koutsaroff, I. P., Eds. 2004; Vol. 784, pp 43-48.
61. Rabe, U.; Janser, K.; Arnold, W., Vibrations of free and surface-coupled atomic force microscope cantilevers: Theory and experiment. *Rev. Sci. Instrum.* **1996,** *67* (9), 3281-3293.
62. Rabe, U.; Arnold, W., ACOUSTIC MICROSCOPY BY ATOMIC-FORCE MICROSCOPY. *Appl. Phys. Lett.* **1994,** *64* (12), 1493-1495.
63. Butt, H. J.; Cappella, B.; Kappl, M., Force measurements with the atomic force microscope: Technique, interpretation and applications. *Surf. Sci. Rep.* **2005,** *59* (1-6), 1-152.
64. Felten, F.; Schneider, G. A.; Saldana, J. M.; Kalinin, S. V., Modeling and measurement of surface displacements in BaTiO3 bulk material in piezoresponse force microscopy. *J. Appl. Phys.* **2004,** *96* (1), 563-568.
65. Shin, J.; Rodriguez, B. J.; Baddorf, A. P.; Thundat, T.; Karapetian, E.; Kachanov, M.; Gruverman, A.; Kalinin, S. V., Simultaneous elastic and electromechanical imaging by scanning probe microscopy: Theory and applications to ferroelectric and biological materials. *Journal of Vacuum Science & Technology B* **2005,** *23* (5), 2102-2108.
66. Karapetian, E.; Kachanov, M.; Kalinin, S. V., Nanoelectromechanics of piezoelectric indentation and applications to scanning probe microscopies of ferroelectric materials. *Philosophical Magazine* **2005,** *85* (10), 1017-1051.
67. Morozovska, A. N.; Eliseev, E. A.; Svechnikov, G. S.; Kalinin, S. V., Nanoscale electromechanics of paraelectric materials with mobile charges: Size effects and nonlinearity of electromechanical response of SrTiO(3) films. *Phys. Rev. B* **2011,** *84* (4), 045402.
68. Kalinin, S. V.; Bonnell, D. A., Temperature dependence of polarization and charge dynamics on the BaTiO3(100) surface by scanning probe microscopy. *Appl. Phys. Lett.* **2001,** *78* (8), 1116-1118.
69. Neumayer, S. M.; Ievlev, A. V.; Collins, L.; Vasudevan, R.; Baghban, M. A.; Ovchinnikova, O.; Jesse, S.; Gallo, K.; Rodriguez, B. J.; Kalinin, S. V., Surface Chemistry Controls Anomalous Ferroelectric Behavior in Lithium Niobate. *ACS Appl. Mater. Interfaces* **2018,** *10* (34), 29153-29160.
70. Garrity, K.; Kolpak, A. M.; Ismail-Beigi, S.; Altman, E. I., Chemistry of Ferroelectric Surfaces. *Adv. Mater.* **2010,** *22* (26-27), 2969-2973.
71. Wang, J. L.; Gaillard, F.; Pancotti, A.; Gautier, B.; Niu, G.; Vilquin, B.; Pillard, V.; Rodrigues, G.; Barrett, N., Chemistry and Atomic Distortion at the Surface of an Epitaxial BaTiO3 Thin Film after Dissociative Adsorption of Water. *Journal of Physical Chemistry C* **2012,** *116* (41), 21802-21809.
72. Garra, J.; Vohs, J. M.; Bonnell, D. A., Defect-mediated adsorption of methanol and carbon dioxide on BaTiO(3)(001). *Journal of Vacuum Science & Technology A* **2009,** *27* (5), L13-L17.
73. Yun, Y.; Kampschulte, L.; Li, M.; Liao, D.; Altman, E. I., Effect of ferroelectric poling on the adsorption of 2-propanol on LiNbO3(0001). *Journal of Physical Chemistry C* **2007,** *111* (37), 13951-13956.
74. Morozovska, A. N.; Eliseev, E. A.; Morozovsky, N. V.; Kalinin, S. V., Ferroionic states in ferroelectric thin films. *Phys. Rev. B* **2017,** *95* (19).

75. Yang, S. M.; Morozovska, A. N.; Kumar, R.; Eliseev, E. A.; Cao, Y.; Mazet, L.; Balke, N.; Jesse, S.; Vasudevan, R. K.; Dubourdieu, C.; Kalinin, S. V., Mixed electrochemical-ferroelectric states in nanoscale ferroelectrics. *Nature Physics* **2017,** *13* (8), 812-818.
76. Jesse, S.; Guo, S.; Kumar, A.; Rodriguez, B. J.; Proksch, R.; Kalinin, S. V., Resolution theory, and static and frequency-dependent cross-talk in piezoresponse force microscopy. *Nanotechnology* **2010,** *21* (40), 405703.
77. Proksch, R.; Kalinin, S. V., Energy dissipation measurements in frequency-modulated scanning probe microscopy. *Nanotechnology* **2010,** *21* (45).
78. Shimizu, S.; Hoyer, P. O.; Hyvarinen, A., Estimation of linear non-Gaussian acyclic models for latent factors. *Neurocomputing* **2009,** *72* (7-9), 2024-2027.
79. Hyvarinen, A.; Zhang, K.; Shimizu, S.; Hoyer, P. O., Estimation of a Structural Vector Autoregression Model Using Non-Gaussianity. *J. Mach. Learn. Res.* **2010,** *11*, 1709-1731.
80. Shimizu, S.; Inazumi, T.; Sogawa, Y.; Hyvarinen, A.; Kawahara, Y.; Washio, T.; Hoyer, P. O.; Bollen, K., DirectLiNGAM: A Direct Method for Learning a Linear Non-Gaussian Structural Equation Model. *J. Mach. Learn. Res.* **2011,** *12*, 1225-1248.
81. Shimizu, S., Joint estimation of linear non-Gaussian acyclic models. *Neurocomputing* **2012,** *81*, 104-107.
82. Shimizu, S.; Bollen, K., Bayesian Estimation of Causal Direction in Acyclic Structural Equation Models with Individual-specific Confounder Variables and Non-Gaussian Distributions. *J. Mach. Learn. Res.* **2014,** *15*, 2629-2652.
83. Hyvarinen, A.; Oja, E., Independent component analysis: algorithms and applications. *Neural Netw.* **2000,** *13* (4-5), 411-430.
84. Hyvarinen, A.; Oja, E., A fast fixed-point algorithm for independent component analysis. *Neural Comput.* **1997,** *9* (7), 1483-1492.
85. Wilson, A. G.; Hu, Z.; Salakhutdinov, R.; Xing, E. P. In *Deep kernel learning*, Artificial intelligence and statistics, PMLR: 2016; pp 370-378.
86. Blei, D. M.; Kucukelbir, A.; McAuliffe, J. D., Variational Inference: A Review for Statisticians. *Journal of the American Statistical Association* **2017,** *112* (518), 859-877.